\documentstyle[prl,aps,epsfig,multicol]{revtex}

\begin{document}

\title{Creep via dynamical functional renormalization group}
\author{P. Chauve{$^{1,2}$}, T. Giamarchi{$^1$} and P. Le Doussal{$^2$}}
\address{{$^1$}Laboratoire de Physique des Solides, CNRS URA 02,
Universit{\'e} Paris-Sud, B{\^a}t. 510, 91405 Orsay, France}
\address{{$^2$}CNRS-Laboratoire de Physique Th{\'e}orique de l'Ecole
Normale Sup{\'e}rieure, 24 rue Lhomond, F-75231 Paris, France}
\maketitle

\begin{abstract}
We study a $D$-dimensional interface driven in a disordered medium. We
derive finite temperature and velocity functional renormalization group
(FRG) equations, valid in a $\epsilon =4-D$ expansion. These equations
allow in principle for a complete study of the the velocity versus
applied force characteristics. We focus here on the creep regime at finite
temperature and small velocity. We show how our FRG approach gives
the form of the $v-f$ characteristics in this regime, and in
particular the creep exponent, obtained previously only through
phenomenological scaling arguments.
\end{abstract}

\begin{multicols}{2}

Lines and surfaces can exhibit a remarkable variety of complex phenomena
when they are driven by an external force over a disordered substrate.
This situation is encountered in numerous physical systems such as
growth phenomena
\cite{krim_growth_review,kardar_review_lines},
magnetic domain wall motion
\cite{lemerle_domainwall_creep},
fluid invasion of porous media
\cite{lenormand_fluid_porous},
vortex systems \cite{blatter_vortex_review},
charge density waves (CDW) \cite{gruner_revue_cdw}
or Wigner crystals
\cite{andrei_wigner_2d}.

In all these cases there is a competition between the elastic energy,
which tends to keep the system straight, and the disorder.
In addition to making the static interface rough, the disorder has also
important consequences for the dynamics. At zero temperature $T=0$, the
interface remains pinned until a critical force $f_c$ is reached,
whereas for large drive, it moves with a velocity proportional to the
force, the disorder being averaged by the fast motion.
Obtaining theoretically the
velocity $v$ versus applied force $f$ characteristics is a
very challenging problem since such a $v-f$ curve is directly
measurable, and is in most cases one of the most important physical
properties ({\it e.g.} the transport properties in vortex, CDW or Wigner
crystal systems).

A very fruitful approach is to cast the depinning transition of an
elastic system in the general framework of critical phenomena, with the
velocity as an order parameter
\cite{fisher_functional_rg,nattermann_stepanow_depinning,%
fisher_depinning_meanfield,narayan_fisher_depinning}. A functional
renormalization group (FRG) treatment of the equation of motion
\cite{fisher_functional_rg,nattermann_stepanow_depinning}
allows to obtain the various critical exponents
characterizing the depinning transition.

At finite temperature, the situation is more complex since the
interface can move by thermal activation below the zero temperature
threshold force. This leads to a rounding of the depinning transition
(see {\it e.g.} \cite{middleton_depinning_rounding}) when $f \sim f_c$, and
to a creep motion of the interface when $f \ll f_c$. Indeed
it was realized that, due to the glassy nature of the \emph{static}
pinned interface, a moving interface would have to overcome divergent
barriers as the applied force is reduced
\cite{feigelman_creep,ioffe_creep,nattermann_rfield_rbond,%
feigelman_collective,nattermann_pinning} in contrast to much older
theories that assumed that motion occured only through finite barriers
\cite{anderson_kim}. This lead to the proposal of a phenomenological
theory of creep resulting in a highly non-linear $v-f$ characteristics
of the form $v \propto \exp \left( -(U_c/T) (f_c/f)^{\mu} \right)$.
Using scaling arguments,
and assuming that relevant barriers for the dynamics scale
with the same exponent as the energies of metastable \emph{static}
configurations, one obtains \cite{feigelman_collective,nattermann_pinning}
$\mu = (D-2+2\zeta)/(2-\zeta)$ where
$D$ is the dimension of the interface and $\zeta$ the \emph{static}
roughening exponent of the interface. This remarkable formula
relates \emph{dynamical} properties to purely static quantities.
The creep theory has been extremely successful in explaining various
physical phenomena, in particular vortex systems
\cite{blatter_vortex_review}, and the quantitative relations between the
exponents has recently been verified by experiments on magnetic domain
walls \cite{lemerle_domainwall_creep}.

\begin{figure}\label{vfins} 
\centerline{\epsfig{file=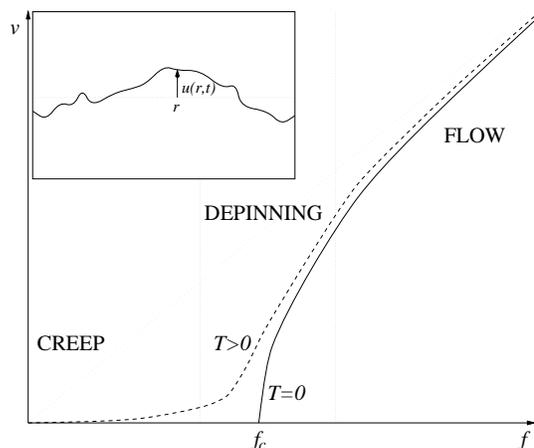,angle=0,width=7cm}}
\caption{\narrowtext
Typical force-velocity curves. Inset: an interface with relative transverse
displacement $u$.}
\end{figure}

Unfortunately, up to now, the creep formula has remained phenomenological
\cite{creep_d1}
and relies on a certain number of mostly unverified
assumptions \cite{creep_d11}. On the other hand,
extending to finite temperatures and velocities the FRG techniques used at
$T=0$ has proved very challenging, although some interesting results
could be obtained for the periodic systems driven within their internal
space \cite{giamarchi_m2s97_vortex,giamarchi_book_young,%
ledoussal_mglass_long,balents_moving_long}.
We address this issue in the present paper. We give a generalization of the
FRG equations valid for arbitrary temperature and velocity. These equations
allow in principle for a complete study of the thermal effects on the $v-f$
characteristics. We show in this letter how they can be used to obtain the
creep behavior and to derive the creep exponent from first principle
({\it i.e.} the equation of motion). We compare our findings with the
phenomenological theory. A more refined study of the flow, a precise
calculation of prefactors, and other consequences will be presented in
a longer publication \cite{chauve_creep_long}.

We consider a $D$-dimensional manifold without overhangs described
by a height function $u$, as shown in Figure~\ref{vfins}, embedded in a
space of dimension $d$. If $D=d$ this also describes periodic systems
\cite{ledoussal_mglass_long}. The system obeys the equation of motion
\begin{equation}
(\eta \partial_t - c \nabla^2)u_{rt}=f-\eta v+F(r,vt+u_{rt})+\zeta_{rt}
\end{equation}
where $\eta$ is the friction coefficient and $c$ is the elastic constant. For
simplicity we assume here isotropic elasticity, but generalization to
less symmetric elastic tensors can easily be done. Thermal
fluctuations are described by the Langevin force
$\langle \zeta_{rt}\zeta_{r't'} \rangle =2\eta T\delta^D(r-r') \delta(t-t')$
(the brackets denote thermal averages).
Disorder gives rise to a random force characterized by
$\overline{F(r,u)F(r',u')}=\delta^D(r-r') \Delta(u-u')$
(the overbar denotes disorder averages).

\begin{figure}\label{distypes}
\centerline{\epsfig{file=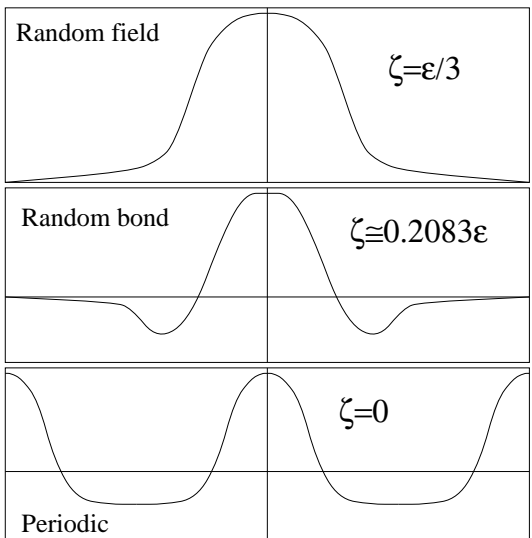,angle=0,width=7cm}}
\caption{\narrowtext Typical correlators for different cases of
disorder and the corresponding static roughening exponent $\zeta$.}
\end{figure}

Depending on the precise microscopic nature of the problem, the correlator
$\Delta$ can have different forms \cite{fisher_functional_rg}
as shown on Figure~\ref{distypes}. The \emph{random field} case occurs when
$\Delta$ is short range, the disorder is \emph{random bond} when the force
is the derivative of a potential which has short range correlations, and
finally, when the system is periodic, such as vortex lattices,
$\Delta$ is periodic \cite{giamarchi_vortex_short}. The total height
of the interface is $vt+u_{rt}$, and the condition $\overline{\langle
\partial_t u_{rt} \rangle}=0$ fixes self-consistently a relation
between the average velocity $v$ and the force $f$. In the absence of
disorder, we are left with the trivial friction law $f=\eta v$. In the
following we denote by $\tilde{f}=f-\eta v$ the deviation from this law.

To make a systematic perturbation expansion and renormalization out of this
stochastic equation, we follow the standard method of dynamic
field theory \cite{janssen_uchap,martin_siggia_rose} which yields an
action $S(u,\hat {u})$, where $\hat {u}$ is the response field.
Within this formalism, a perturbative expansion in disorder $\Delta$ allows
to generalize \cite{ledoussal_mglass_long} the large velocity expansion
\cite{larkin_largev,schmidt_hauger} of the characteristics $f(v)$ to
finite temperatures. But the main interest of this formalism is to
allow to go beyond perturbation theory and to build the FRG equations
at finite temperatures and velocities.
The procedure is similar to the one used in
\cite{ledoussal_mglass_long}, so we just give an outline here and
leave technical details for \cite{chauve_creep_long}.
We expand the weight
$e^{-S}$ up to second order in $\Delta$, knowing that
$\tilde{f}={\mathcal O} ( \Delta )$. We then integrate over the ``fast''
modes of the fields $u$ and $\hat {u}$ having Fourier components
$\Lambda e^{-l}<q<\Lambda$, where $1/\Lambda$ is a short distance
cutoff. Reexponentiation gives an effective action similar to
the original one, but with renormalized parameters, once we restore the
original cutoff by rescaling the fields as
$u'(r,t)=e^{-\zeta l}u(e^l r,e^{zl} t)$ and
$\hat{u}'(r,t)=e^{-\sigma l} \hat{u}(e^l r,e^{zl} t)$.
The elasticity $c$ is not renormalized, for statistical tilt symmetry reasons
\cite{nattermann_stepanow_depinning,narayan_fisher_depinning}.
This fixes $\sigma$ such that the trivial scaling of $c$ is zero.
We are then left with two free parameters (which can be chosen
$l$-dependent if needed): $\zeta$, the roughening exponent of the
interface relating longitudinal and transverse lengths $u \sim
r^\zeta$, and $z$, the dynamical exponent relating space and time
$t\sim r^z$. Denoting $\frac{S_D\Lambda^D}{(c\Lambda^2)^2}\Delta$ by
$\Delta$, $\frac{S_D\Lambda^D}{c\Lambda^2}T$ by $T$, $\frac{\eta v}{c
\Lambda^2}$ by $\lambda$, keeping $\Lambda$ fixed ($S_D$ is the
surface of the sphere divided by $(2\pi)^D$) and performing the
rescalings, we obtain the following flow equations in a $\epsilon=4-D$
expansion
\begin{eqnarray}
\partial_l  \Delta(u) &=&(\epsilon-2\zeta)\Delta(u)+\zeta u
\Delta'(u) + T \Delta''(u) \nonumber \\&&
+\int_{s>0,s'>0}\!\!\!\!\!\!\!\!\!\!\!\!\!\!\!\!\!\!\!\!e^{-s-s'}
\left[ \Delta''(u)\left(
\Delta((s'-s)\lambda)-\Delta(u+(s'-s)\lambda) \right)\right.\nonumber \\
&&- \Delta'(u-s'\lambda)\Delta'(u+s\lambda) \nonumber \\
&&\left.+\Delta'((s'+s)\lambda)\left(
\Delta'(u-s'\lambda)-\Delta'(u+s\lambda) \right) \right]
\label{equ:disorderrg}
\end{eqnarray}
\begin{eqnarray}
\partial_l  \ln \lambda&=&2-\zeta-\int_{s>0}e^{-s}s \Delta''(s\lambda)\\
\partial_l  \ln T&=&\epsilon-2-2\zeta+\int_{s>0}e^{-s}s \lambda
\Delta'''(s\lambda)\\
\partial_l  \tilde{f}&=&(2-\zeta)\tilde{f}+c \Lambda^2 \int_{s>0} e^{-s}
\Delta'(s\lambda)\\
\partial_l  c&=&0\\
\partial_l  \ln v&=&z-\zeta   \label{equ:vrg}
\end{eqnarray}
These equations \cite{similar_equations} allow in principle for a
complete description of the properties of the interface at finite
temperatures and velocities.

For the deterministic ($T=0$) static ($v=0$) fixed point, the flow equations
are greatly simplified since $T_l \equiv 0$ and $\lambda _l \equiv 0$.
One recovers the usual formulas derived in the context of the depinning
transition via replicas
\cite{fisher_functional_rg,giamarchi_vortex_short} or FRG
\cite{nattermann_stepanow_depinning,narayan_fisher_depinning}.
The correlator $\Delta_l$ flows to a fixed point $\Delta^*$ which depends
on the type of disorder. Exact solutions are known  for random field
\cite{fisher_functional_rg} and periodic cases
\cite{giamarchi_vortex_short}, and
asymptotics for random bond. In all these cases, the fixed
point function has a cusp ($\Delta^{*\prime}$ is discontinuous at the
origin), which appears at a finite scale
$l_c=\frac{1}{\epsilon }\log (1+\frac{\epsilon }{-3 \Delta ''(0)})$
and which is responsible for the existence of the pinning force
\cite{fisher_functional_rg,giamarchi_vortex_short}.
The scale at which the divergence of $\Delta ''(0)$ occurs is the Larkin
length, typical size of a segment of the interface wandering over the
correlation length of the disorder. Its expression $e^{l_c}/\Lambda$
coincides with its standard \cite{larkin_70} expression,
$R_c \simeq (\epsilon \frac{c^2 r_f^2}{\Delta (0)})^{1/\epsilon}$,
when restoring the original $\Delta$, and denoting its range by $r_f$. In the
following, we denote by $\alpha$ the length $-\Delta^{*\prime}(0+)$,
where \cite{chauve_creep_long}
$\alpha \simeq 0.4 \epsilon^{2/3} \left(\int \! \Delta _{l=0} \right)^{1/3}$
for the random field case and
$\alpha = \epsilon a /6$
for the periodic case where $a$ is the lattice spacing.

The flow equations (\ref{equ:disorderrg}-\ref{equ:vrg}) at finite but
small velocity and zero temperature provide a natural derivation of
the procedure used in
\cite{nattermann_stepanow_depinning} for calculating the critical
force and the depinning exponents. Since up to the Larkin scale,
disorder and displacements are small, the flow is trivial.  After
$l_c$, $\Delta $ has reached its fixed point $\Delta ^*$.  Provided
that the velocity is small ($\lambda_l $ is smaller than the range
$r_f^*$ of $\Delta ^*$), our flow equations coincide with those used in
\cite{nattermann_stepanow_depinning} since we can replace the r.h.s. of the
flow equations by the values at $0+$ of $\Delta^*$.
The scale used to cut the flow in \cite{nattermann_stepanow_depinning}
appears here naturally when the length $\lambda_l$ is of the
order of the range $r_f^*$. As in
\cite{nattermann_stepanow_depinning}, we denote by $l_V$ the
corresponding value of $l$. Note that $r_f^*$ can be quite different
from the original range $r_f$ of the disorder. At larger scales, one
crosses over for $l \sim l_V$ to a regime where the disorder acts as
Langevin forces, and we are left with an effective Edwards Wilkinson
equation of motion. We thus recover \cite{chauve_creep_long}
\begin{equation}\label{equ:critforce}
f_c=-\frac{c\Lambda ^2 \Delta ^{*\prime}(0+)}{2-\zeta}e^{(\zeta-2)l_c}
\end{equation}
This expression gives back, up to an $\epsilon$ coefficient, the
dimensional estimate $f_c \simeq \frac{c
r_f}{R_c^2}$, since $R_c$ is $e^{l_c}/\Lambda$ and $\alpha e^{\zeta
l_c}\simeq 0.2 \epsilon r_f$ in the random field case and $\alpha e^{\zeta
l_c} = \epsilon a/6$ in the periodic case.

\begin{figure}\label{cusp}
\centerline{\epsfig{file=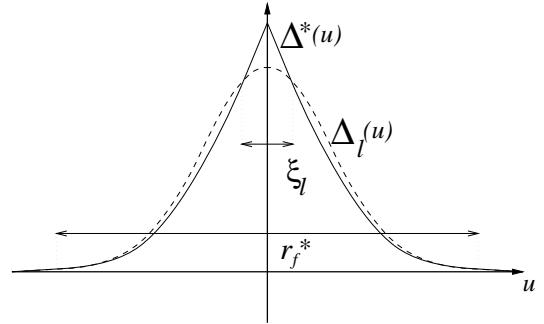,angle=0,width=7cm}}
\caption{\narrowtext Generation of the cusp: two lengthscales appear.
the width $\xi_l$ of the rounding, of order $T_l$, and the width
$r_f^*$ of the fixed point function $\Delta^*$.}
\end{figure}

Let us now analyse the finite temperature case. A finite temperature
flows to zero with the free energy exponent $\epsilon-2-2\zeta$.
But the presence of the $T \Delta''(u)$ term in (\ref{equ:disorderrg})
is enough to smooth the behavior of the correlator near the origin,
and it removes the divergence of $\Delta''(0)$.
Indeed one expects $\Delta_l$ to remain analytic
at all \emph{finite} scales even if its limit $\Delta^*$ has a cusp.
More precisely, during the flow, $-\Delta_l''(0)$ increases and one may
assume \cite{fisher_conjecture} that $-T_l \Delta_l''(0)$
converges, since the flow equation at the
origin is $\partial_l \Delta_l(0)=(\epsilon-2\zeta)\Delta_l(0)
+T_l \Delta_l''(0)$. One can check that its limit is $\alpha^2$
(with $\alpha$ defined above). Using this property in the $T>0$, $v=0$
FRG equations for $\Delta $ and $T$ we obtain the precise way
\cite{chauve_creep_long}
the successive derivatives of $\Delta $ grow with $l$, and get
\begin{equation}\label{equ:cuspform}
\frac{1}{T_l }\left(\Delta _l (0)-\Delta _l (\frac{T_l
}{\alpha}x) \right) \longrightarrow \sqrt{1+x^2}-1
\end{equation}
The rounding close to the origin appears at a scale of the order of the
temperature $T_l $. The form of $\Delta _l $ in this regime is
given by $\Delta^*$
for $u\gg \xi_l = \frac{T_l}{\alpha}$ and by the rounding
(\ref{equ:cuspform}) for $u\ll \xi_l$, as shown on Figure~\ref{cusp}.

This property of the finite temperature equations enables us to
investigate the creep regime. In addition to the Larkin scale $l_c$,
two obvious lengthscales appear in the equations: the scale $l_T$
for which $\lambda_l = \xi_l$, the width of the rounding of
$\Delta_l$, and as before, $l_V$ for which $\lambda_l$ is of the
order of the range $r_f^*$ of $\Delta^*$, and above which disorder is
washed out. Provided $f \ll f_c$, the velocity is
arbitrarily small, and so is the initial value of $\lambda$. One has
thus $l_c < l_T < l_V$. The equation for the temperature introduces
in fact an additional lengthscale that we discuss later.
Ignoring for the time being this additional lengthscale allows to
distinguish two main regimes above the Larkin scale $l_c$. In the
thermal regime $l_c < l < l_T$, we can replace the set of equations
(\ref{equ:disorderrg}-\ref{equ:vrg}) by static ones obtained
by replacing the r.h.s. occurences of $\lambda _l$ by zero, \emph{and not}
by $0+$ as for the depinning case. The quantity $-T_l \Delta_l''(0)$
becomes constant \emph{until} $\lambda _l\sim \xi_l$.
In this regime $\lambda$ grows rapidly. Physically this corresponds to a
regime where barriers grow and the system can only move through thermal
activation over those barriers.

Beyond $l_T$ one enters the \emph{depinning} regime. The
disorder is close to its fixed point and the other quantities now flow with
the depinning flow equations, since the integrals over $s$ do no more see the
behaviour \emph{at} the origin but the completely different values of
$\Delta^*$ at $0+$. Roughly speaking we can in this regime ignore the
temperature, as far as motion is concerned. Physically this is because
above the length $e^{l_T}/ \Lambda \sim R_c(f_c/f)^{1/(2-\zeta)}$, that
corresponds \cite{chauve_creep_long}
to the criterion $\lambda_l \sim \xi_l$, the external
force is dominant over the pinning force, and these lengthscales do not
need thermal activation to move. For $l \gg l_T$, the
temperature is not important for the motion. As for
$T=0$, we cut the flow at the scale $l_V$ where a perturbative
expansion in disorder is well defined. We can use it to
estimate the value of $\tilde{f}_{l_V}$. Since the renormalized
temperature is very small it is enough to do this calculation as if
$T_{l_V}=0$. Then it becomes obvious that $\tilde{f}_{l_V}$ does not
depend on the initial velocity, force or temperature.

Solving the flow (\ref{equ:disorderrg}-\ref{equ:vrg}) for the three
regimes yields a relation between initial
quantities. Using (\ref{equ:critforce}) and replacing the \emph{initial}
value of $\tilde{f}$ by $f$ since in the creep regime $f\gg \eta v$, we get
\begin{equation} \label{equ:creepint}
\frac{\eta v}{f_c} = \frac{(2-\zeta)S_D}{(e^{l_c}/\Lambda)^{D-2}
(\alpha e^{l_c})^2}\frac{T}{c}
(\frac{f}{f_c})^{1+\mu}
\exp \left( -\frac{U_c}{T}((\frac{f_c}{f})^{\mu}-1)\right)
\end{equation}
where $\mu=(2-\epsilon+2\zeta)/(2-\zeta)$ and a rough estimate of $U_c$
gives
\begin{equation}\label{equ:barrierrf}
U_c=\frac{c (\alpha e^{\zeta l_c})^2
(e^{l_c}/\Lambda)^{2-\epsilon}}
{(2-\epsilon+2\zeta)S_D} =
\frac{c (C \epsilon r_f)^2 R_c^{2-\epsilon}}
{(2-\epsilon+2\zeta)S_D}
\end{equation}
where $C$ is a constant.
We thus recover precisely the creep exponent $\mu$, obtained previously
only via the phenomenological barrier arguments. Our derivation allows
to obtain this value directly from the equation of motion without having
to do any supplementary assumption. The fact that we do recover the
simple scaling exponent $\mu$, is thus also a strong indication that
the barriers important for the motion do scale with the free energy
exponent (up to subleading corrections). Although our calculation is
not expected to give, at that level, a precise estimate of the
prefactors in the barriers $U_c$ it is conforting to also recover an
expression similar to the one  derived by scaling arguments, $U_c
\simeq c r_f^2 R_c^{D-2}$, {\it i.e.} the barriers at the lengthscale of the
Larkin length. As we showed above the FRG also identifies correctly
the relevant lengthscales appearing in the simple scaling approach of
the creep. This simple application of the FRG equations then gives a
consistent picture for the creep.

The main interest of the present method is of course to allow in
principle to go beyond the simple scaling. In this respect
various questions remain. In particular the FRG equation for the
temperature leads to an additional lengthscale at $\lambda_l \sim
T_l^{3/2}$. Above this lengthscale the disorder starts renormalizing the
temperauture upwards. Although this effect has no impact on the creep
exponent \cite{chauve_creep_long} it can obviously lead to a
modification of the barriers (or equivalently of the $1/T$ term in
(\ref{equ:creepint})). A more precise calculation of this amplitude,
taking into account the full crossover around $l \sim l_T$ will
be presented in \cite{chauve_creep_long}. Given the success of
the finite temperature FRG for the creep regime, it is also tempting to
apply these methods to the
study of other regimes of the motion. In particular, close to the threshold at
finite temperature, a scaling approach describing the motion as a succession
of avalanches, and a complementary FRG approach based on the flow equations
derived here, should help fixing the form of the force-velocity curve.


\begin{thebibliography}{10}

\bibitem{krim_growth_review}
J. Krim and G. Palasantzas, Int. J. Mod. Phys. B {\bf 9},  599  (1995).

\bibitem{kardar_review_lines}
M. Kardar cond-mat/9704172 and cond-mat/9507019.

\bibitem{lemerle_domainwall_creep}
S. Lemerle {\it et~al.}, Phys. Rev. Lett. {\bf 80},  849  (1998).

\bibitem{lenormand_fluid_porous}
R. Lenormand, J. Phys. C {\bf 2},  SA79  (1990).

\bibitem{blatter_vortex_review}
G. Blatter {\it et~al.}, Rev. Mod. Phys. {\bf 66},  1125  (1994).

\bibitem{gruner_revue_cdw}
G. Gr{\"u}ner, Rev. Mod. Phys. {\bf 60},  1129  (1988).

\bibitem{andrei_wigner_2d}
E.~Y. Andrei and {al.}, Phys. Rev. Lett. {\bf 60},  2765  (1988).

\bibitem{fisher_functional_rg}
D.~S. Fisher, Phys. Rev. Lett. {\bf 56},  1964  (1986).

\bibitem{nattermann_stepanow_depinning}
T. Nattermann, S. Stepanow, L.~H. Tang, and H. Leschhorn, J. Phys. (Paris) {\bf
  2},  1483  (1992).

\bibitem{fisher_depinning_meanfield}
D.~S. Fisher, Phys. Rev. B {\bf 31},  1396  (1985).

\bibitem{narayan_fisher_depinning}
O. Narayan and D.~S. Fisher, Phys. Rev. B {\bf 48},  7030  (1993).

\bibitem{middleton_depinning_rounding}
A.~A. Middleton, Phys. Rev. B {\bf 45},  9465  (1992).

\bibitem{feigelman_creep}
M. Feigelman, Sov. Phys. JETP {\bf 58},  1076  (1983).

\bibitem{ioffe_creep}
L.~B. Ioffe, V.~M. Vinokur, J. Phys. C {\bf 20},  6149  (1987).

\bibitem{nattermann_rfield_rbond}
T. Nattermann, Europhys. Lett. {\bf 4},  1241  (1987).

\bibitem{feigelman_collective}
M. Feigelman, V.~B. Geshkenbein, A.~I. Larkin, and V.~M. Vinokur,
Phys. Rev. Lett. {\bf 63},  2303  (1989).

\bibitem{nattermann_pinning}
T. Nattermann, Phys. Rev. Lett. {\bf 64},  2454  (1990).

\bibitem{anderson_kim}
P.~W. Anderson and Y.~B. Kim, Rev. Mod. Phys. {\bf 36},  39  (1964).

\bibitem{creep_d1}
In $d=1$, a creep formula was derived for a particle moving in a correlated
  potential. See P. Le Doussal and V.~M. Vinokur Physica C {\bf 254} 63 (1995).

\bibitem{creep_d11}
Some progres could be made in $d=1+1$ where it was found that, up to
  logarithmic factors, the barriers did scale as assumed. See e.g. L. V.
  Mikheev, B. Drossel and M. Kardar Phys. Rev. Lett. {\bf 75} 1170 (1995).

\bibitem{giamarchi_m2s97_vortex}
T. Giamarchi and P. {Le Doussal}, Physica C {\bf 282-287},  363  (1997).

\bibitem{giamarchi_book_young}
T. Giamarchi and P. {Le Doussal},  in {\em Statics and dynamics of disordered
  elastic systems}, edited by A.~P. Young (World Scientific, Singapore, 1998),
  p.\ 321, cond-mat/9705096.

\bibitem{ledoussal_mglass_long}
P. {Le Doussal} and T. Giamarchi, 1997, cond-mat 9708085; to appear in Phys.
  Rev. B (Mai 1998).

\bibitem{balents_moving_long}
L. Balents, M.~C. Marchetti, and L. Radzihovsky, Phys. Rev. B {\bf 57},  7705
  (1998).

\bibitem{chauve_creep_long}
P. Chauve, T. Giamarchi and P. {Le Doussal}, to be published.

\bibitem{giamarchi_vortex_short}
T. Giamarchi and P. {Le Doussal}, Phys. Rev. Lett. {\bf 72},  1530  (1994).

\bibitem{janssen_uchap}
H.~K. Janssen, Z. Phys. B {\bf 23},  377  (1976).

\bibitem{martin_siggia_rose}
P.~C. Martin, E.~D. Siggia, and H.~A. Rose, Phys. Rev. A {\bf 8},  423  (1973).

\bibitem{larkin_largev}
A.~I. Larkin and Y.~N. Ovchinnikov, Sov. Phys. JETP {\bf 38},  854  (1974).

\bibitem{schmidt_hauger}
A. Schmidt and W. Hauger, J. Low Temp. Phys {\bf 11},  667  (1973).

\bibitem{similar_equations}
See also \cite{ledoussal_mglass_long} and L. H. Tang and S. Stepanow
  (unpublished). The last reference did not obtain the first equation giving
  the renormalisation of the disorder at finite temperature and velocity.

\bibitem{larkin_70}
A.~I. Larkin, Sov. Phys. JETP {\bf 31},  784  (1970).

\bibitem{fisher_conjecture}
D.~S. Fisher (unpublished).

\end{thebibliography}

\end{multicols}
\end{document}